\documentclass[aip,jcp,reprint,twocolumn,amsmath,amssymb]{revtex4-1} 
\pdfoutput=1

\usepackage{amsmath,amsthm}
\usepackage{graphicx}
\usepackage{dcolumn}
\usepackage{bm}
\usepackage{color}
\usepackage{epsfig}
\usepackage{multirow}
\usepackage{siunitx}
\usepackage[caption=false]{subfig}

\usepackage[disable]{todonotes}
\makeatletter
\presetkeys
    {todonotes}
    {inline}{}
\makeatother

\graphicspath{{./figures/}{./sketches/}}

\newcommand{\Vect}[1]{\ensuremath{\bm{#1}}} 

\newcommand{\es}{ESPResSo}

\newcommand{\etal}{et~al.~}

\newenvironment{changed}{\begingroup}{\endgroup}

\begin{document}

\preprint{Preprint}

\title{The influence of charged-induced variations in the local permittivity on the static and dynamic properties of polyelectrolyte solutions}

\author{Florian \surname{Fahrenberger}}
\author{Owen A. \surname{Hickey}}
\author{Jens \surname{Smiatek}}
\author{Christian \surname{Holm}} \email{holm@icp.uni-stuttgart.de}

\affiliation{Institute for Computational Physics, University of Stuttgart, Stuttgart 70569, Germany}


\date{\today}

\begin{abstract}
There is a large body of literature investigating the static and dynamic properties of polyelectrolytes due both to their widespread application in industrial processes and their ubiquitous presence in biology. Because of their highly charged nature, polyelectrolytes tend to alter the local dielectric permittivity of the solution within a few nanometers of their backbone. This effect has, however, been almost entirely ignored in both simulations and theoretical work. In this article we apply our recently developed electrostatic solver based on Maxwell's equations to examine the effects of the permittivity reduction in the vicinity of the polyelectrolyte. 
We first verify our new approach by calculating and comparing ion distributions around a linear fixed polyelectrolyte and find both quantitative and qualitative changes in the ion distribution. Further simulations with an applied electric field show that the reduction in the local dielectric constant increases the mobility of the chains by approximately ten percent. More importantly, variations in the local dielectric constant lead to qualitatively different behavior of the conductivity.
\end{abstract}

\maketitle

\section{Introduction}

Electrophoresis is the directed motion of an object in an aqueous solution subject to an external electric field. Electric fields are often used for the characterization, purification, and manipulation of polymers~\cite{viovy00a,dorfman10a,shendruk12a}, colloids~\cite{russel89a,wiersema66a,obrien78a,lobaskin07a}, and cells~\cite{hayashi01a,duval10a}, which all tend to ionize in aqueous solutions. While it is tempting to understand electrophoresis as the balance between the external electric force and hydrodynamic friction, the physics is somewhat more complicated~\cite{ohshima06b}. The reason for this is the aggregation of oppositely charges ions, termed counterions, near the surface of the object. The ions closest to the surface make up the Stern layer, strongly adsorbed ions which co-move with the electrophoresing object, effectively reducing the charge of the object. Ions further away from the surface form the diffuse layer, which has a characteristic size given by the Debye length, $\lambda_\text{D}$:
\begin{equation}
    \lambda_\text{D} = \left( 4 \pi \lambda_\text{B} \sum_{j=1}^{N} n_j
    z^2_j \right )^{-1/2},
\label{eq:debye}
\end{equation}
where $N$ is the total number of charge species, $n_j$ and $z_j$ are the number concentration and valency of species $j$. The Bjerrum length $\lambda_\text{B}$ is given by:
\begin{equation}
 \lambda_\text{B}=e^2/(4\pi\varepsilon_0\varepsilon_\text{rs}k_\text{B}T),
\end{equation}
where $\varepsilon_0$ and $\varepsilon_\text{rs}$ are the vacuum and the relative permittivity, respectively.

The ions in the diffuse layer move in the opposite direction of the analyte since they are oppositely charged. The counterions reduce the velocity of the electrophoresing object, through hydrodynamic and electrostatic coupling to the analyte. The plane separating these two layers is called the shear plane, and the two layers are often called electrical double layer (EDL).


Simulations of polyelectrolyte electrophoresis have, for the most part, focused on the electrophoresis of a single chain in bulk solution (free solution electrophoresis). The first results in this regard were lattice-Boltzmann simulations by Grass and Holm~\cite{grass08a,grass08c,grass09a,grass10a}, and multi-particle collision dynamics simulations by Frank and Winkler~\cite{frank08a,frank09a}. Both methods were able to reproduce the sharp rise in electrophoretic mobility as a function of the polymer length for short chains, followed by a slight decrease for longer chains, as is seen in experiment~\cite{hoagland99a}. In contrast, when long-ranged hydrodynamic interactions were ignored, the simulations showed a monotonic decrease in the mobility in stark contrast to experiment. The reason for the monotonic decline in mobility in the absence of hydrodynamic interactions is that a fraction of the counterions strongly bond to the polymer backbone, co-moving with the chain, and effectively reducing the linear charge density~\cite{grass09a,frank09a,hickey12a}. This phenomenon, termed counterion condensation, was first proposed by Manning~\cite{manning69a}. 

A sharp increase in the mobility with increasing chain length is observed for short chains. This increase can be understood through the cooperative shearing of the fluid within the Debye layer~\cite{shendruk12a}. Because of this, the mobility starts to flatten out when the chain length becomes comparable to the Debye length~\cite{grass09a}. The aforementioned condensation of counterions and subsequent reduction of the effective charge of the chain is responsible for the slight decrease in mobility for long chains~\cite{grass09a,frank09a,hickey12a}.

A number of studies have also looked at the electrophoresis of polyelectrolytes through simple channels. Smiatek and Schmid investigate the role of surface slip on the mobility of a polyelectrolyte electrophoresing between parallel plates~\cite{smiatek10a,smiatek11a}. More recently, Yan et~al looked at the electric field dependent mobility of a polyelectrolyte between parallel plates~\cite{yan12a}. The electrophoretic motion of chains through entropic traps~\cite{han00a} has also been studied extensively using computer simulations~\cite{duonghong08a,duonghong08b,fayad10a,laachi07a}. A number of studies have also looked at the electrophoresis of DNA in post arrays~\cite{trahan10a,cho10a,dorfman10a}.

Electric fields are also often applied to polyelectrolyte solutions in order to measure their conductivity and characterize the properties of the solution~\cite{cametti14a,dobryin95a,liparostir09a,colby97a,kwak75a,manning69a,schmitt73a,bordi04a,bordi05a}. Theory is able to correctly predict the conductivity under a variety of solvent conditions and salt concentrations~\cite{cametti14a,dobryin95a,manning69a}. The conductivity normalized by the concentration of charge carriers is called the equivalent conductivity and is often used instead of the raw conductivity. Deviations from a simple linear scaling of the conductivity with respect to the concentration, due to both electrostatic and hydrodynamic interactions, causes the variations in the equivalent conductivity. In the salt-free case, there are a number of experiments which show an initial decrease in the equivalent conductivity, followed by an increase starting at polyelectrolyte concentrations of approximately $c\approx 0.01M$~\cite{liparostir09a,colby97a,kwak75a}. Scaling theories~\cite{colby97a,dobryin95a} are able to correctly predict the initial decay, but fail to explain the increase in the equivalent conductivity at polyelectrolyte concentrations above $c\approx 0.01M$~\cite{colby97a}. This increase in the equivalent conductivity appears to be independent of the type of polyelectrolyte used~\cite{liparostir09a,colby97a,kwak75a}, the length of the polyelectrolytes~\cite{colby97a}, and the temperature of the solvent~\cite{liparostir09a}.

In most theories and simulations on charged systems the presence of water molecules is taken into account as a constant dielectric background. This is most often done by setting a constant relative permittivity of the system $\varepsilon_\text{r}=78.5$, corresponding to salt-free water at $10\,^{\circ}{\rm C}$. However, it has been shown that the presence of charged salt ions~\cite{hess06a,linse08a} and charged surfaces~\cite{bonthuis11a} can considerably alter the local dielectric constant.
Recent theoretical work has shown that taking into account the smoothly changing dielectric properties between a spherical colloid and the surrounding fluid causes a depletion in counterions near the surface~\cite{fahrenberger13c,ma14a,nakamura13a,curtis15a}. This effectively increases the thickness of the Debye layer and in turn reduces the hydrodynamic and electrostatic coupling of the surface to its counterions.
Presumably, a similar effect occurs for polyelectrolytes, though the magnitude of the effect is, a priori, unclear. Only a few techniques exist for calculating ion distributions in the presence of a varying dielectric constant~\cite{tyagi10a,jadhao12a,jadhao13a}, and for the most part these methods have only been used for calculating static ion distributions.

The influence of a varying dielectric background consists of three separate contributions: The dielectrically reduced pairwise electrostatic interaction between charges can be described by the aforementioned algorithms, as well as the influence of dielectric enclosures on a point charge by means of virtual boundary or image charges~\cite{tyagi10a,neumann83c,lindell93a,messina02f,boda04b,tyagi07a,tyagi08a,qin09a,arnold13c,kesselheim11a}, or via a functional approach~\cite{jadhao12a,jadhao13a,bichoutskaia10a}. However, in addition to these two contributions, mobile ions have self energy fluctuations in an inhomogeneous dielectric medium. In uniform dielectric media, the solvation energy of charged particles is invariant and can be absorbed into the chemical potential. However, this term cannot be disregarded when the solvent has a varying dielectric permittivity. As pointed out in an earlier publication~\cite{fahrenberger13c}, this direct Coulomb term can be described by a modification of the Born energy~\cite{born20a} by introducing the local self energy contribution. The self energy of ion $i$ is given by
\begin{equation}
U_i^\text{self}= \frac{q_i^2}{4\pi\varepsilon_0} \left[ \frac{G_\text{pol}(\Vect{r}_i,\Vect{r}_i)}{2}
+\frac{1}{2\varepsilon(\Vect{r}_i)a_i} \right],
\label{eq:solvation-energy}
\end{equation}
where the first term including $G_\text{pol}$ is the contribution from virtual surface or image charges, and the second term relates to the solvation energy within the medium. Here, $q_i$ is the charge and $a_i$ is the Born radius of the ion. If the gradient of the locally varying permittivity $\varepsilon(\Vect{r})$ is non-zero, this energy leads to a force that is included in the HIM method presented in our earlier work with Xu~\cite{fahrenberger13c} and the electrostatic algorithm used in this manuscript.

In this article, we expand on our previous study on the conductivity of polyelectrolyte solutions with a locally varying permittivity~\cite{fahrenberger15b}. We first introduce the new electrostatic algorithm in detail. The algorithm extends a lattice-based electrostatic algorithm based on Maxwell's equations. We first examine the case of a line of fixed, charged monomers and compared the results to Poisson-Boltzmann theory around a cylinder. We then apply the algorithm to a free polyelectrolyte in aqueous solution and examine the influence of the resulting permittivity gradients on the couterion distribution and radius of gyration. Next, we look at how the variations in the local dielectric constant affect the electrophoretic mobility of polyelectrolytes and the conductivity of polyelectrolyte solutions. We end with a brief conclusion.

\section{Simulation Method}
\subsection{Molecular Dynamics}
We perform Molecular Dynamics (MD) simulations using the Extensible Simulation Package for Research on Soft matter, \es{}~\cite{limbach06a,arnold13a,misc-espressomd}.
There is a purely repulsive Weeks-Chandler-Anderson (WCA) potential~\cite{weeks71a} between all particles to represent steric interactions:
\begin{equation}
    \label{eq:wca}
    V_{\textrm{WCA}}(r)=\left\{
      \begin{array}{l l}
          4\varepsilon_\text{MD}\left(\left(\frac{\sigma_\text{LJ}}{r}\right)^{12} -
          \left(\frac{\sigma_\text{LJ}}{r}\right)^6 + \frac{1}{4}\right), & \quad
          r < 2^{1/6}\sigma_\text{LJ},\\
          0, & \quad \text{otherwise} ,
      \end{array} \right.
\end{equation}
where $r$ is the distance between the particles, $\sigma_\text{LJ}=\SI{0.3}{nm}$ is the fundamental MD length scale, and $\varepsilon_\text{MD}=k_\text{B}T=4.11\times 10^{-21}J$ is the fundamental MD energy scale.

$M_\text{p}$ polyelectrolytes consist of a linear chain of negatively charged monomers $q_1=-1e$ (where $e$ is the fundamental MD unit of charge) of length $N$. Adjacent monomers are linked together using a finitely-extensible nonlinear elastic (FENE) potential
\begin{equation}
	V_{\textrm{FENE}}(r)=-\frac{kR_0^2}{2}\ln\left(1-\left(\frac{r}{R_0}\right)^2\right),
\end{equation}
where $R_0=1.5\sigma_\text{LJ}$ is the maximum extension of the bond and $k=30\varepsilon_\text{MD}/\sigma_\text{LJ}^2$ is the energy scale of the bond. An equal number $M_\text{p}N$ of counterions with charge $q=1e$ are added to the box to keep the net charge of the system zero. All particles have a mass $m=1m_0$, where $m_0$ is the fundamental MD unit of mass. Velocities and positions are updated using the velocity Verlet algorithm with a time step $\Delta t_\text{MD} = 0.01 \tau$, where $\tau=\sqrt{m_0\sigma_\text{LJ}/\epsilon}$ is the MD unit of time. In simulations where we measure static properties, we use a Langevin thermostat with a temperature $k_\text{B}T=\varepsilon_\text{MD}$ and a friction constant $\Gamma_\text{Langevin}=1m_0/\tau$. All of our simulations used periodic boundary conditions in all directions.

\subsection{Hydrodynamic Interactions}
\label{sec:HI}
In simulations where we calculate dynamic properties, we chose the lattice-Boltzmann (LB) method to model the hydrodynamic interactions, since it is extremely efficient and has been shown to produce physically sound results~\cite{lobaskin07a,grass08a,raafatnia14b,raafatnia15a}. Specifically, we use ESPResSo's D3Q19 lattice-Boltzmann (LB)~\cite{rohm12a} with a kinematic viscosity $\nu=0.8\sigma_\text{LJ}^2/\tau$ and density $\rho_\text{fluid}=1 m_0/\sigma_\text{LJ}^3$. The LB grid spacing is $a_\text{LB}=\SI{0.4}{nm}$ and the time step was set to $\Delta t_\text{LB} = 0.01 \tau$. These parameters will give the correct hydrodynamic radius of $0.44a \approx \SI{0.15}{nm}$ for counterions while keeping the lattice-Boltzmann algorithm stable. We applied an external electric field of $E_\text{ext}=0.1\varepsilon_\text{MD}/\sigma_\text{LJ} e$ (which has been shown to be low enough to still be in the linear response regime~\cite{lobaskin04b,grass09a}) when measuring mobilities and conductivities. The temperature was held constant using an LB thermostat with a temperature $k_\text{B}T=\varepsilon_\text{MD}$.
\subsection{Electrostatics}
Electrostatic algorithms in molecular dynamics typically deal with variations in the dielectric constant using a functional, boundary charges, or mirror charges~\cite{jadhao12a,tyagi10a,neumann83c,linse86a,lindell93a,messina02f,boda04b,tyagi07a,tyagi08a,linse08a,rescic08a,qin09a,arnold13c}. These methods have a number of disadvantages for studying polyelectrolytes. Most importantly, such methods do not allow for gradual changes in the dielectric constant, but only sharp discontinuities in the permittivity. In addition, there is often no way to have charges move across a gradient in the dielectric permittivity, or to change the background bulk permittivity with time.
Our algorithm~\cite{fahrenberger14a} does not have these limitations. It is based on the local electrodynamic solver introduced by Maggs and Rossetto in 2002~\cite{maggs02a}. The algorithm was later adapted for off-lattice Molecular Dynamics simulations by Rottler and Maggs~\cite{rottler04a, rottler04b}, and in parallel by Pasichnyk and D\"unweg~\cite{pasichnyk04a}. 

Traditional approaches typically calculate the electrostatic potential $\Phi$ by solving the static Poisson equation
\begin{equation}
	\nabla \varepsilon \nabla \Phi = -\rho ,
	\label{eq:poisson}
\end{equation}
where $\varepsilon$ is the dielectric permittivity, and $\rho$ is the charge density. Our algorithm instead focuses on solving the electrodynamic equivalent, Gauss' law
\begin{equation}
	\nabla \Vect{D} = \rho ,
	\label{eq:gauss-law}
\end{equation}
where $\Vect{D} = \varepsilon \Vect{E}$ includes the dielectric permittivity and the local electric field $\Vect{E}$. This works because electrostatics is merely the limit of electrodynamics where the speed of light $c$ approaches infinity. The wave propagation speed $c$ can be reduced significantly and the algorithm still reproduces the same particle dynamics and thermodynamic observables~\cite{rottler04a,rottler04b,pasichnyk04a}. This is very similar to Car-Parrinello Molecular Dynamics~\cite{car85a}, where the velocity of the electrons is reduced to unrealistic values, yet the dynamics of the corresponding nuclei remains accurate. In our electrostatic algorithm, the reduction of the propagation speed makes the method computationally feasible and has been shown to require a similar amount of computation time compared to more established methods~\cite{arnold13b}. The use of electrodynamics for solving electrostatic interactions in molecular dynamics simulations has been coined Maxwell Equations Molecular Dynamics (MEMD).

We have already implemented several other algorithms that allow for variations in the local dielectric constant within the \es{} software package~\cite{fahrenberger14a,arnold13a,misc-espressomd}. The method method presented here nevertheless offers a few key advantages compared to other electrostatics algorithms. Most importantly, the method is intrinsically local, and allows for an arbitrary spatial and temporal variation of the local dielectric permittivity. We have successfully applied this implementation to a colloid with a smooth radial change of the dielectric permittivity in the vicinity of the colloid surface~\cite{fahrenberger13c}. We showed that a region of varying dielectric permittivity, into which charges can enter, gives rise to a strong force in the direction of the permittivity gradient and therefore significantly influences the structure of the electric double layer (EDL) around the colloid~\cite{fahrenberger13c}. This force is the result of a finite sized object in a permittivity gradient, that results from the self-energy. This self-energy term is missing from methods which only include a discrete jump in the dielectric constant and significantly affects the equilibrium distribution of ions. We also compared our implementation to a Monte Carlo code that can deal with radially symmetric and smooth permittivity gradients and found excellent agreement~\cite{fahrenberger13c}, giving us confidence on the reliability of the algorithm for spatially varying dielectric permittivity. The MEMD grid spacing was set to $a_\text{MEMD}=\SI{0.4}{nm}$, the time step was $\Delta t_\text{MEMD} = \Delta t_\text{MD}$, and the artificial mass was $f_\text{mass} = 0.05 m_0$.


\section{The influence of charges on dielectric permittivity}

In MD simulations of polyelectrolytes, an implicit water model is often used since explicit water molecules would increase the necessary compute time by more than an order of magnitude~\cite{slater09a}. To get reliable results in the case of electrophoresis, the implicit water model needs to correctly reproduce two properties of water: the hydrodynamic interactions (which we deal with using lattice-Boltzmann as explained in section~\ref{sec:HI}) and the screening of electrostatic interactions. Most MD simulations deal with the screening of electrostatics by introducing a bulk dielectric permittivity~$\varepsilon$. This shows up as a constant prefactor in the Poisson equation~\eqref{eq:poisson}, and since it does not depend on any local parameters, can simply be included in the electrostatic force calculation using the same prefactor.

This electrostatic screening effect originates in the rearrangement of water molecule dipoles in the presence of an electric field~$\Vect{E}$. On average, the dipoles will have a tendency to align with the electric field lines and create an additional field in the opposite direction. The resulting electric field~$\Vect{E}(\Vect{r})$, at a position~$\Vect{r}$, will be weakened by a factor of~$\varepsilon(\Vect{r})$, which leads to the definition and use of the displacement field~$\Vect{D}=\varepsilon(\Vect{r})\Vect{E}$. As implied by the notation~$\varepsilon(\Vect{r})$, this prefactor is not necessarily constant throughout the system but can depend on the local surroundings. Specifically, if there are charged particles like polyelectrolyte monomers or salt ions present, the water molecules in the vicinity of these charges will be restricted in their rotational degrees of freedom by the strong electric field create by the charge particle. The water molecules thus cannot react to external electrostatic influences as freely as bulk water molecules. This means that the dielectric response in the presence of salt ions is reduced. Several studies have examined the dependence of the bulk permittivity on the salt concentration in water~\cite{lamm97a,hess06a,hess06b,gavryushov06a}, as well as the variation in the permittivity around a charged object~\cite{bonthuis11a}.

Most current electrostatics algorithms only allow for a change in the system-wide dielectric permittivity, and this parameter can be adjusted in polyelectrolyte simulations in an attempt to account for the salt concentration in the solution, an approach we also investigate in this manuscript. However, it is apparent that the permittivity is not constant in a system that includes a highly charged object surrounded by counterions, and a bulk phase that contains mostly solvent and relatively few co- and counterions. This is particularly true for a dilute polyelectrolyte solution. The local ion concentration will not only vary in the proximity of the polyelectrolyte, but also in time since the polyelectrolyte is mobile. In the present article, we will introduce a method to deal with these spatial and temporal changes in the local dielectric constant and examine their influence on the static and dynamic properties of polyelectrolyte solutions.

\section{Iterative approach}
\label{sec:iterative}

\begin{figure}[!htb]
	\includegraphics[width=0.9\linewidth]{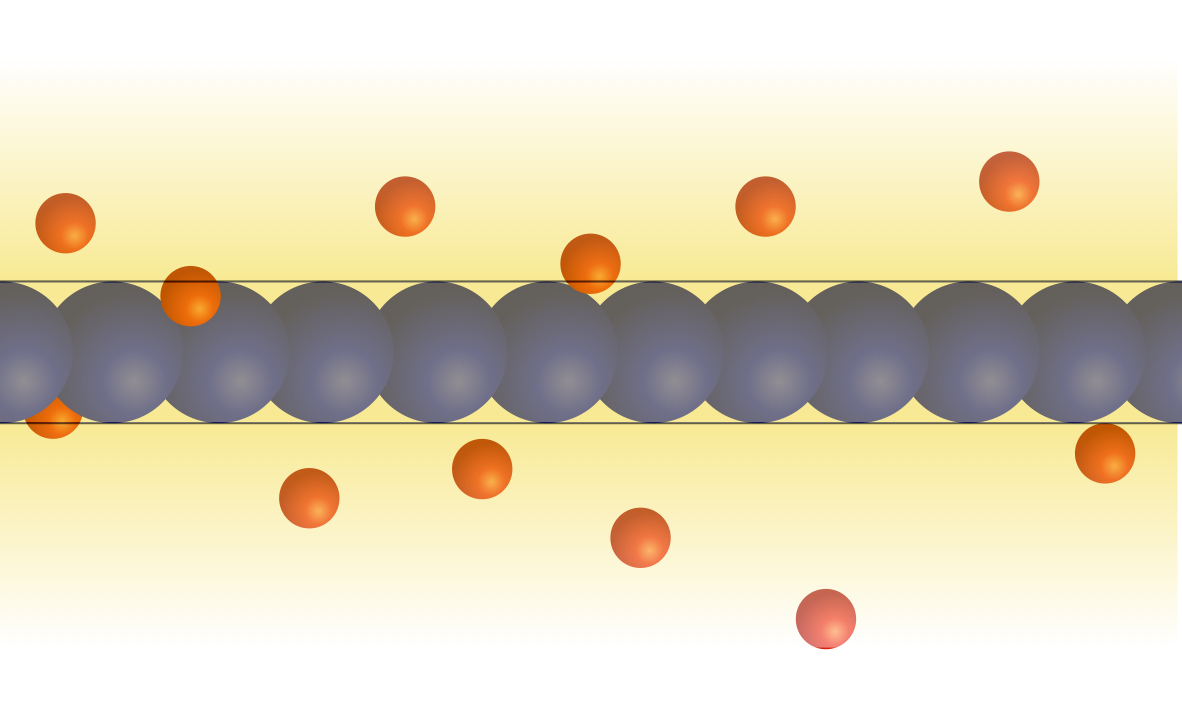}
	\caption{Schematic of our first simulation setup: a fixed line of monomers (blue) is placed in the center to form a stiff rod and counterions (red) can move freely around the polyelectrolyte. The permittivity (yellow background) is set to $\varepsilon = 2$ within the rod and is variable outside the rod, eventually reaching $\varepsilon = 78.5$ (white) at longer distances.}
	\label{fig:rod-scheme}
\end{figure}

As a starting point, we consider a polyelectrolyte to be a charged rod, or in our case a linear arrangement of charged beads fixed in space as shown in Fig.~\ref{fig:rod-scheme}. The rod consists of $N=80$ monomers which are fixed in place for the duration of the simulation and spaced $\SI{0.3}{nm}$ apart in a cubic box with a side length of $\SI{24}{nm}$ with periodic boundary conditions.
Inside the cylinder representing the polyelectrolyte, there are very few water molecules and the dielectric permittivity is set to $\varepsilon = 2$\cite{pethig85a,dobrynin05a}. Outside the polyelectrolyte, we expect the dielectric permittivity to depend on the local concentration of counterions, and gradually increases to around $\varepsilon = 78.5$ in the bulk.

If we average the salt concentration over time, we should get a cylindrical ion profile around the rod center because of the system's symmetry. This amounts to a one-dimensional problem to solve, and there are certainly faster and more efficient ways to approach it than an MD simulation. However, we wanted to introduce a verification for our algorithm, and we wanted a point of reference for later comparison when the dielectric constant varies both spatially and temporally. To this end, we will solve for the electrostatic forces using the implementation of MEMD with varying dielectric permittivity in \es{}.

To calculate the local permittivity value from the local ion concentration, we require a function that maps a given salt concentration to a permittivity value. For our simulations, we use the empirical function obtained by Hess et~al.~\cite{hess06a} from atomistic MD simulations of a sodium chloride salt solution:
\begin{equation}
	\varepsilon = \frac{78.5}{1+0.278\cdot C} \quad ,
	\label{eq:salt-map}
\end{equation}
where $C$ is the salt concentration in moles per liter [M].

\begin{figure}[!htb]
	\includegraphics[width=0.9\linewidth]{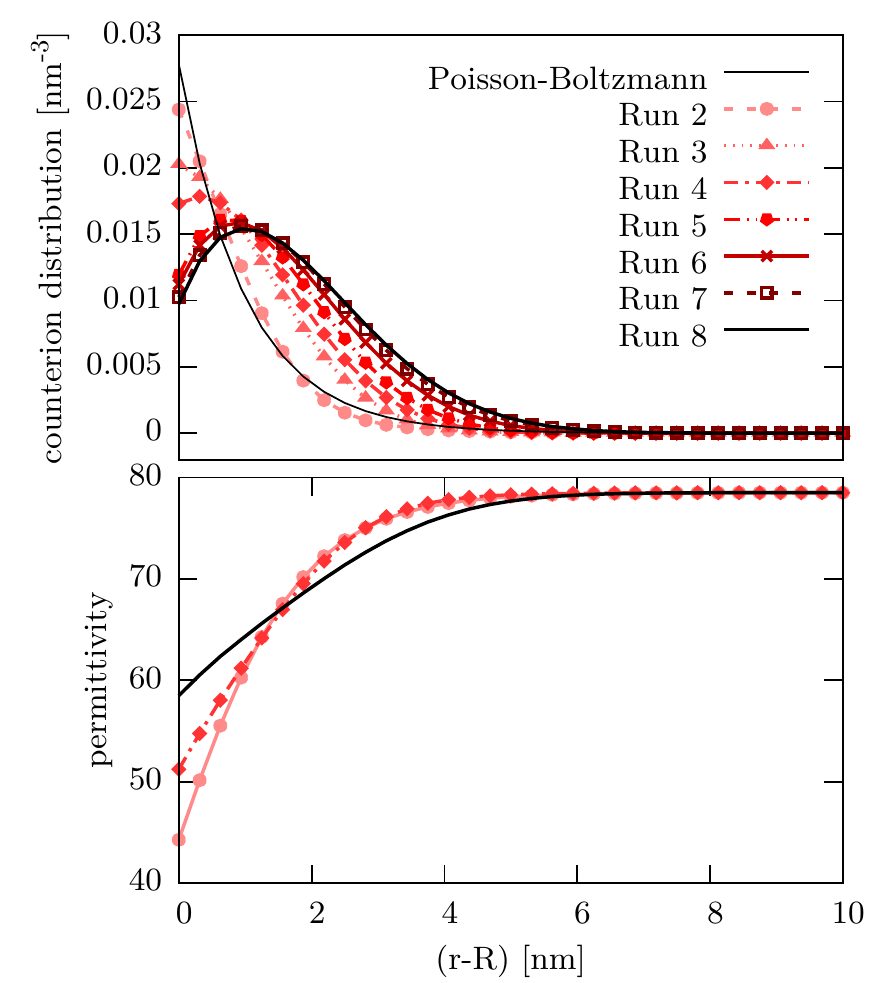}
	\caption{Starting from a flat permittivity profile of $\varepsilon=78.5$, the counterion distribution including a salt concentration dependent permittivity around a charged stiff rod is calculated iteratively. The counterion distribution converges to a stable solution (Run 8, top graph) which is significantly different from the often applied Poisson-Boltzmann solution (thin black line, top graph). The first iterative run expectedly gives a result identical to the Poisson-Boltzmann distribution. The dielectric permittivity, calculated from equation~\eqref{eq:salt-map} including an additional contribution from the polyelectrolyte charge, shows a weaker attraction between the counterions and the rod (bottom graph).}
	\label{fig:iterations-stiff-rod}
\end{figure}

We self-consistently solve for the counterion distribution and the dielectric constant using an iterative scheme. We start out with a flat permittivity profile, which assumes a constant $\varepsilon = 78.5$ for the surrounding solvent. We then map this salt concentration to a dielectric permittivity via equation~\eqref{eq:salt-map}, and interpolate the resulting permittivity distribution to a grid. We then set the permittivity values of the lattice links and, after equilibration, run the simulation until the average radial counterion distribution converged. We use a successive under-relaxation, where the two preceding ion distributions are averaged to stabilize the iterative scheme. The averaged ion distribution is then used to generate new permittivity values and the process is repeated iteratively until both the ion distribution and permittivity converge. The results of the first eight runs are presented in Fig.~\ref{fig:iterations-stiff-rod}.

The simulations clearly converge towards a stable final distribution, and for the simulation shown in Fig.~\ref{fig:iterations-stiff-rod} it is converged after the eighth iteration. The converged counterion density actually increases in the first two nanometers, before an exponential decrease from two nanometers to around five nanometers from the rod. This roughly corresponds to the region of reduced dielectric permittivity shown in the bottom graph of Fig.~\ref{fig:iterations-stiff-rod}, which also extends approximately five nanometers from the rod. \begin{changed}It is counterintuitive at first glance that a decreased permittivity close to the polyelectrolyte would lead to a decrease in counterions at the surface, since the electrostatic attraction increases. However, the effect seen here is mostly due to the force that the permittivity gradient $\nabla\varepsilon$ exerts on a counterion, which pushes the charges away from the backbone, along the permittivity gradient.\end{changed}

The counterion distribution in Fig.~\ref{fig:iterations-stiff-rod} looks similar to our findings for spherical colloids~\cite{fahrenberger13c}, which is to be expected given that the underlying physics is independent of the specific geometry. Because of these similarities and the very stable final distribution, we are confident that our simulation method produces a realistic counterion distribution for a stiff rod model including varying permittivity effects. However, this method is time consuming and not suitable for dynamic simulations since the dielectric constant does not vary temporally. Nonetheless, it does provide a good point of comparison for the adaptive approach introduced in the next section.

\section{Adaptive approach}

To gain a more flexible and widely applicable algorithm, we opted to calculate the instantaneous local salt concentration during the simulation. We again simulate a stiff rod consisting of $N=80$ monomers fixed in space with a distance $\SI{0.3}{nm}$ between monomers, in a cubic box with a side length of $\SI{24}{nm}$ with periodic boundary conditions.

\begin{figure}[!htb]
	\includegraphics[width=0.6\linewidth]{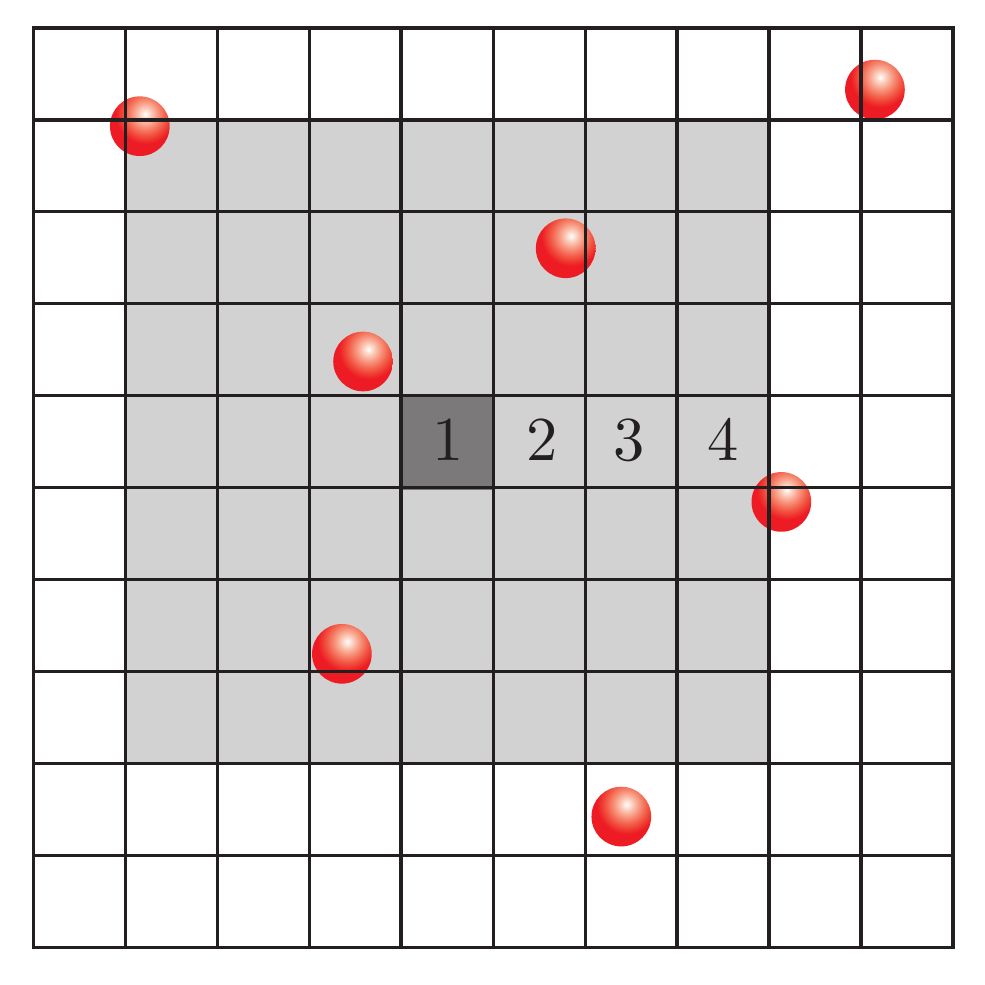}
	\caption{To calculate a local salt concentration on the fly, we use a weighted average in the surrounding $7^3$ lattice cells. The weights are equal to the inverse of the shell number, meaning the central cell gets a weight of 1, the second shell a weight of $1/2^2$, the third shell a weight of $1/3^2$ and the outermost cells getting a weight of $1/4^2$. 
	This calculation is carried out for every MEMD cell individually.                                                                                                                                                                                                                                                                                        
The result can then be mapped to a dielectric permittivity for the center cell using equation~\eqref{eq:salt-map}.}
	\label{fig:grid-averaging}
\end{figure}

For every lattice cell we calculate a weighted average of the ion concentration within a predefined cube, as shown in Fig.~\ref{fig:grid-averaging}. \begin{changed}The ion concentration is calculated for every single MEMD cell in the simulation box and treats charged monomers exactly the same as the salt ions. This means that the local ion concentration is simply a metric for how many charged particles are in the vicinity of the MEMD cell.\end{changed} If the mesh is chosen too coarsely or the screening length too small, there will be large jumps in the permittivity when ions enter or leave the cube. This should be avoided since dielectric jumps are physically unrealistic and represent barriers that block ion flow almost entirely. For our system, we have chosen a cube size of $7^3$ lattice cells. Given the fine mesh of the simulation, this corresponds to a reach of $\SI{1.4}{nm}$ or $2l_B$ (Bjerrum lengths) in water, which is a realistic screening length and we found it to result in sufficiently smooth permittivity curves. The charges included in the surrounding lattice cells are weighted by the inverse square of the shell number (see Fig.~\ref{fig:grid-averaging}). This represents a $1/r^2$ influence, in accordance with the assumption that the polarization response is linear to the electric field, which also decays as $1/r^2$.

\begin{figure}[!htb]
	\includegraphics[width=0.9\linewidth]{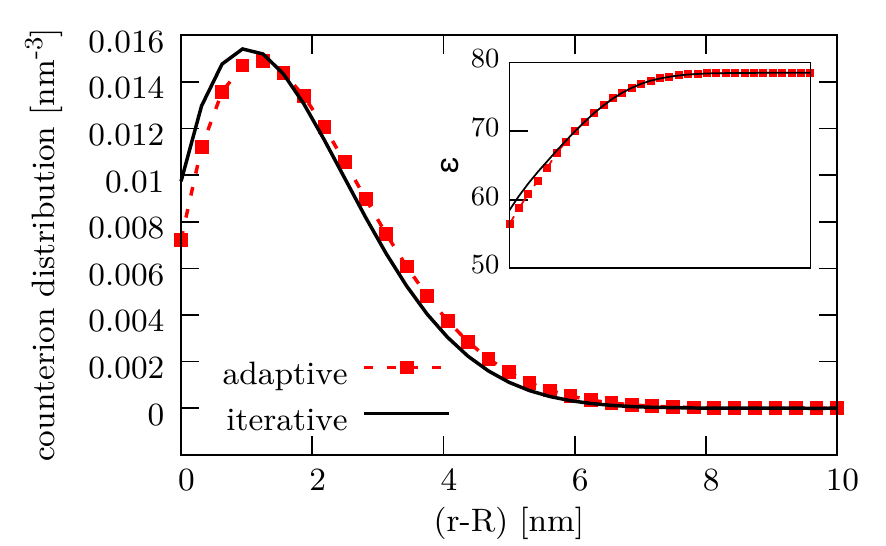}
	\caption[]{Comparison of the newly adopted scheme to the final distribution reached with our iterative approach. \begin{changed}The inset shows the permittivity distribution around the rod. Both\end{changed} results are almost identical.}
	\label{fig:adaptive-scheme-rod}
\end{figure}

For verification of this new adaptive scheme to calculate the local charge concentration, we again simulated the system sketched in Fig.~\ref{fig:rod-scheme}. We did not manually set the dielectric permittivity within the polyelectrolyte and ran the simulation starting with a random distribution of counterions. The result and comparison to the iterative approach \begin{changed}for the counterion distribution and the permittivity\end{changed} is shown in Fig.~\ref{fig:adaptive-scheme-rod}.

The solution provided by our new adaptive scheme results in almost the exact same structure of the electric double layer. Instead of a monotonic exponential decrease, the counterions are pushed away from the polyelectrolyte and reach a stable maximum at around $\SI{1.1}{nm}$ from the surface. This shows that the counterion distribution is not sensitive to the exact method used to calculate the local dielectric constant, which indicates that our somewhat ad hoc method of calculating the local ion concentration correctly captures the underlying physics.

\section{Flexible Polyelectrolyte}

With this new adaptive scheme, we are now able to fully simulate a flexible polyelectrolyte in aqueous solution with the local dielectric constant dependent on the local salt concentration. The simulation is still computationally demanding, since the MEMD lattice has to be very fine to accurately represent the changes in dielectric permittivity around the polyelectrolyte, and because of the calculation of the salt concentration for every cell using the surrounding $7^3$ neighbor lattice sites. It is, however, far less costly than simulations with explicit water molecules. Depending on the number of processors, it runs about a factor of $2$ to $3.5$ slower than simulation with constant background permittivity.

\begin{figure}[htb]
	\includegraphics[width=0.8\linewidth]{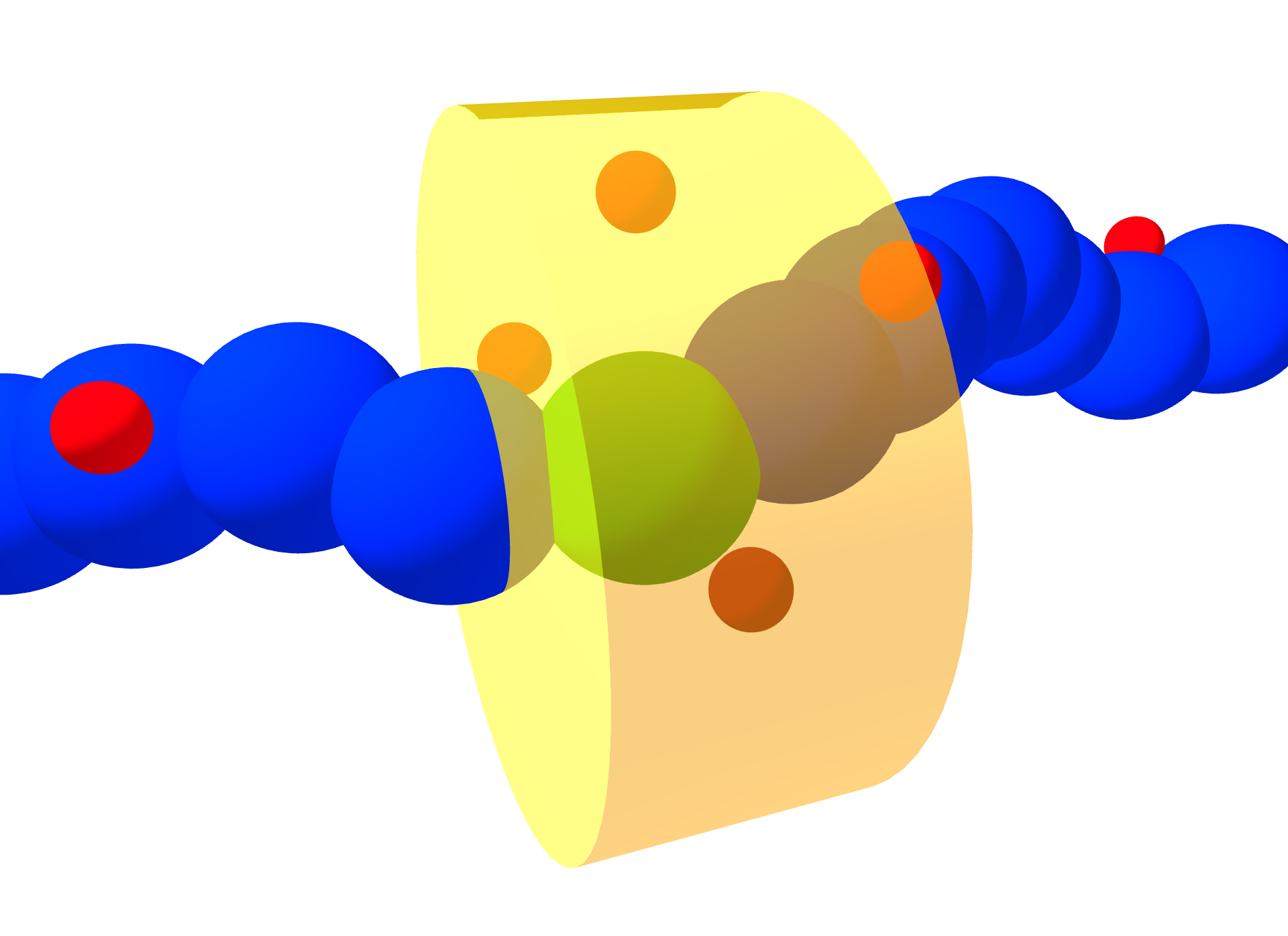}
	\caption{To calculate the radial distribution function of counterions for a flexible polyelectrolyte, we form a cylinder around each monomer, the axis of which is parallel to the connection between the two direct neighbor monomers.}
	\label{fig:flexible-rdf-scheme}
\end{figure}

For a flexible polymer, it is not as straight forward to obtain a radial distribution function of counterions. However, for better comparison we adopted the calculation presented in Fig.~\ref{fig:flexible-rdf-scheme}.

For each monomer, we calculate the vector between the centers of the two neighboring monomers to which it is bonded. This vector is then shifted to the monomer's center. Finally, we calculate the radial distribution function in a cylinder of length $2d$, where $d$ is the equilibrium distance between two monomers.

\begin{changed}
It should be pointed out that with this scheme, it is possible for a single counterion to be included in the distribution function of several different monomers at the same time step. To ensure a valid distribution function, it is therefore imperative to keep a running total of the total number of counterions counted throughout the simulation and normalize the distribution function accordingly.
\end{changed}

\begin{figure}[!htb]
	\subfloat[comparison of the three methods\label{fig:RDF-comparison}]{%
		\includegraphics[width=0.9\linewidth]{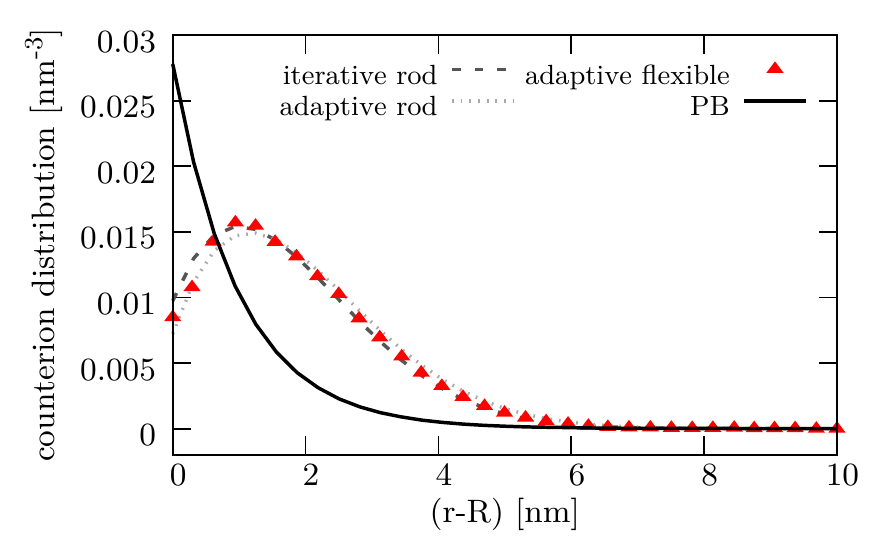}
	}\\
		\subfloat[atomistic simulations\label{fig:RDF-atomistic}]{%
		\includegraphics[width=0.9\linewidth]{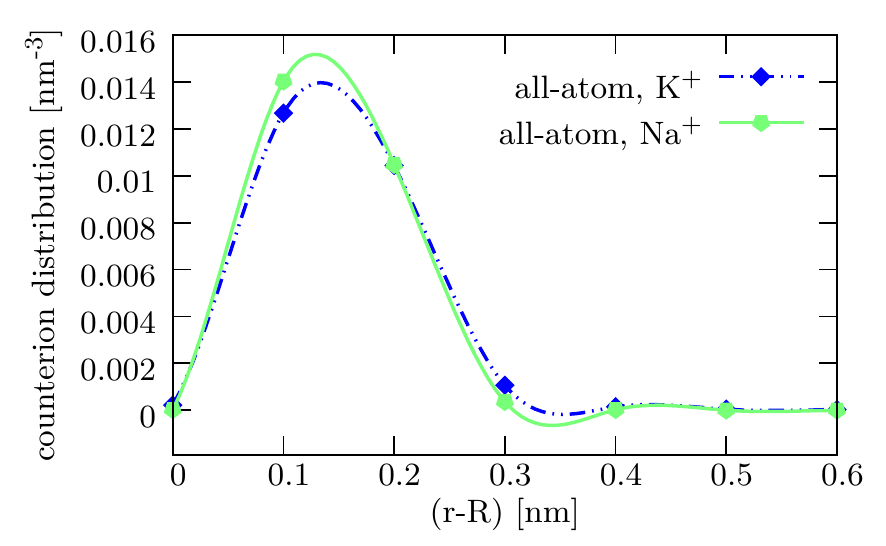}
	}
	\caption[]{Comparison of the counterion distribution functions for a flexible polyelectrolyte (red triangles), a stiff rod using the iterative (dashed line) and adaptive (dotted line) schemes, and the Poisson-Boltzmann solution that does not take smoothly varying permittivity into account (solid line). The depletion of counterions close to the backbone is present in atomistic simulations as well. \begin{changed}The initial rise in the counterion concentration and thickening of the Debye layer is due to a force pushing the counterions away from the polyelectrolyte backbone. This force is proportional to the gradient of the dielectric constant and physically is the result of the solvation energy of an ion being higher in a dielectric background with a larger dielectric constant.\end{changed}}
	\label{fig:RDFs}
\end{figure}

To verify that the scheme works with a flexible polyelectrolyte, we simulated a single chain with $50$ monomers, each carrying a charge of $q_m = -1e$. To keep the system neutral we added 50 counterions with charge $q_{ci}=1e$ but no additional salt ions. We again use a periodic box with side length $\SI{24}{nm}$. A comparison of the result for this flexible polyelectrolyte to the schemes applied to the stiff rod, to a simple Poisson-Boltzmann solution, and to atomistic simulations are presented in Fig.~\ref{fig:RDFs}. The three models that include a salt dependent variation of the dielectric permittivity show quite good agreement. They all differ \emph{qualitatively} from the Poisson-Boltzmann solution.
Note that we previously showed that simulations with a sharp dielectric jump turns out to be almost identical to the Poisson-Boltzmann result~\cite{fahrenberger15b}.

An additional confirmation is that our simulations qualitatively match atomistic simulations, as seen in Fig.~\ref{fig:RDF-atomistic}. The all-atom Molecular Dynamics simulations were performed with the GROMACS 4.5.5 software package~\cite{pronk13a} at $\SI{300}{K}$ for a model polyelectrolyte in aqueous solution and in the presence of counterions. The system studied is a sodium chloride (NaCl) solution with a Kirkwood-Buff based force field~\cite{gee11a} in combination with the SPC/E water model~\cite{berendsen87a}. Kirkwood-Buff force fields have been shown to reproduce thermodynamic and static properties in good agreement to experimental findings and to avoid the spurious artifacts of other force fields, like the overestimation of ion pairing effects. We constructed a simple fictitious linear polyelectrolyte with 30 'CH2' beads as defined in the GROMOS force field~\cite{schuler01a} with the corresponding monomeric binding distance of $\SI{0.149}{nm}$. The polyelectrolyte can be interpreted as a rod of infinite length by using the periodicity of the simulation box with a cubic side length of $\SI{3.80473}{nm}$ in agreement to the approach presented in Ref.~\cite{heyda12a}. We assigned a charge of $q=\pm 1e$ to every second monomer while the other monomers remain uncharged, which gives a line charge density of $l= \pm \SI{3.94}{e/nm}$. The Bjerrum length $\lambda_\text{B}=\SI{0.78}{nm}$ for the SPC/E water model~\cite{heyda12a} yields a Manning parameter $\zeta=l\lambda_\text{B}/e=3.08 \gg 1$ which indicates a large fraction of condensed counterions. We randomly inserted $N_c=15$ counterions in the box to achieve electroneutrality, which gives a salt concentration of $c=\SI{0.45}{M}$. Finally, the positions of the monomers were fixed which avoids the influence of configuration effects on the ion distribution~\cite{smiatek14a}.
The largest difference between the coarse-grained and atomistic simulation in Fig.~\ref{fig:RDFs} is that the counterion density decays much faster in the case of the atomistic simulations. This is because the density of counterions in the box is much higher, which effectively decreases the Debye length $\lambda_\text{D}$~\cite{tessier06a}.

\subsection{Radius of Gyration and Diffusion of an Isolated Polyelectrolyte}

\begin{figure}[!htb]
	\includegraphics[width=0.9\linewidth]{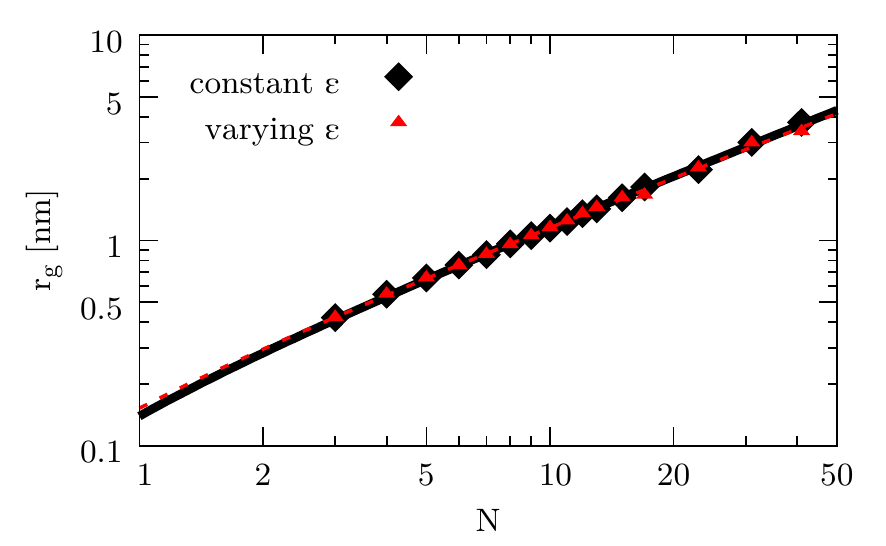}
	\caption{The radius of gyration as a function of the chain length with a uniform permittivity (black diamonds) and with varying permittivity (red triangles). We have included exponential fit functions that give a scaling behavior of $0.79$ and $0.78$ for simulations with and without varying permittivity, respectively.}
	\label{fig:radii-of-gyration}
\end{figure}

Despite the relatively large difference in the counterion distribution around the chain when variations in the local dielectric constant are taken into account, most static properties of the polyelectrolyte, like the radii of gyration or end-to-end distance, were surprisingly unaffected by including the local drop of the dielectric constant in the vicinity of the polyelectrolyte. This is clearly seen in the radius of gyration values plotted in Fig.~\ref{fig:radii-of-gyration}. Note that here we scaled the box length with the cube root of the chain length $L=27 N^{1/3}$, in order to keep the counterion density constant, and thus the Debye length.
Both with and without taking into account variation in the local permittivity constant, the radius of gyration scales as $N^{0.8}$, in good agreement with previous work~\cite{grass08c,frank09a,hickey14a}. 

\begin{figure}[!htb]
	\includegraphics[width=0.9\linewidth]{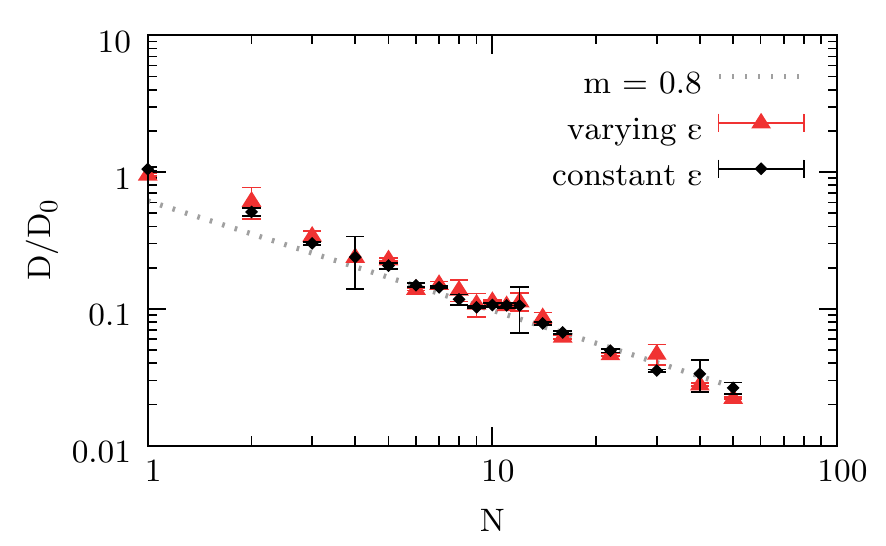}
	\caption{Normalized diffusion of a flexible polyelectrolyte with and without varying dielectric permittivity. Both simulations follow the predicted $D=D_0x^{-m}$ behavior. The scaling parameter changes slightly from $m=0.93(\pm 0.03)$ for varying to $m=1.03(\pm 0.02)$ for constant permittivity. However, this is mainly due to deviations from the expected $m=0.8$ behavior at very short polymer lengths, as can be seen by comparing with the dotted line.}
	\label{fig:diffusion}
\end{figure}

The diffusion of the polyelectrolyte, as shown in Fig.~\ref{fig:diffusion}, leads to slightly different parameters when fitting the power law
\begin{equation}
	D = D_0 x^{-m}
\end{equation}
to the data, where the fit parameter $D_0$ represents the diffusion coefficient of a single isolated monomer. The exponent decreases from $m=1.03(\pm 0.02)$ for constant permittivity to $m=0.93(\pm 0.03)$ for varying permittivity. This is somewhat surprising since based on Zimm's theory we would expect that the diffusion coefficient would scale with the same exponent as the radius of gyration. The fact that the diffusion coefficient drops off slightly faster is partly due to the rather small size of the periodic box, which creates large finite-size corrections. The main contribution is, however,
that the scaling law is only strictly valid in the limit of long chains.
Fig.~\ref{fig:diffusion} shows that the scaling exponent $0.8$, obtained from the radius of gyration data, can also reasonably be used to fit the data at large values of $N$.

\subsection{Electrophoresis}


\begin{figure}[!htb]
	\includegraphics[width=0.9\linewidth]{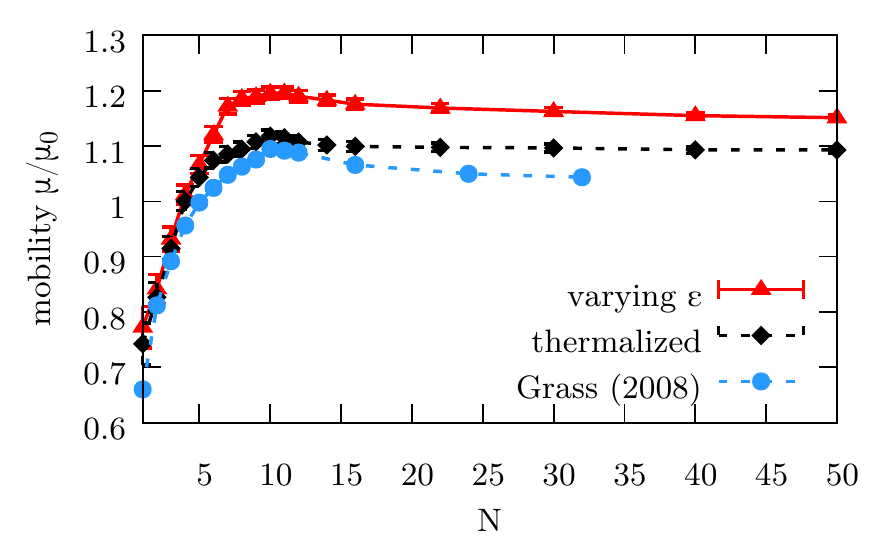}
	\caption{Normalized electrophoretic mobility of the polyelectrolyte for constant and varying dielectric background. While both data sets show a similar qualitative behavior, the mobility is significantly increased when variations in the local permittivity are taken into account. For comparison, and to judge the influence of ghost mode thermalization, we have included the results obtained by Grass~\etal~\cite{grass08a}.}
	\label{fig:mobility}
\end{figure}
In this section, we compare the data of Grass and Holm~\cite{grass08a} including hydrodynamic interactions to the same simulation with a Lattice-Boltzmann algorithm including thermalized ghost modes and finally to a new simulation featuring locally and temporally varying permittivity, as calculated by our adaptive approach. Our simulations were carried out using a single polymer of lengths $N$, with one charge $q_m=-1$ per monomer. $N$ counterions with a charge $q_{ci}=1e$ were also added to the system, but no additional salt. The box size was set to $L=27 N^{1/3} \sigma_\text{LJ}$ to maintain a constant counterion density (which also keeps the effective Debye length constant~\cite{tessier06a}) for all simulations. We measured the velocity of the polymers in the direction of the field and divided by the magnitude of the external field to get the mobility values in Fig.~\ref{fig:mobility}. The mobility is normalized using $\mu_\text{red} = \mu/\mu_0$ with $\mu_0 = e/6\pi\eta l_B$.
The inclusion of spatially and temporally varying permittivity shows the same qualitative behavior, but produces mobilities which are significantly larger than the simulations with a uniform dielectric constant. 

The mobility values can be interpreted as the ratio of an effective charge to that of an effective friction coefficient~\cite{grass08a,frank09a,hickey12a,shendruk12a}
\begin{equation}
	\mu =  \frac{Q_\text{eff}}{\Gamma_\text{eff}},
\end{equation}
where $Q_\text{eff}$ and $\Gamma_\text{eff}$ are the effective charge and effective friction of the chain respectively.
We can calculate the effective charge per monomer of the polyelectrolyte three different ways: (i)~Dynamically, by measuring the mobility in a Langevin dynamics simulation in response to an externally applied electric field. (ii)~Statically, by determining the integrated charge shift of the moving inflection point according to the ion distribution. And (iii)~we can calculate a theoretical prediction with the assumption of Manning condensation,

According to Manning's theory, the predicted value for condensed counterions for a Poisson-Boltzmann distribution is
\begin{equation}
	N_\text{cci} = 1 - \frac{\sigma_\text{bond}{l_B}} \approx 1 - \frac{\SI{0.243}{nm}}{\SI{0.714}{nm}} \approx 1 - 0.34 ,
	\label{eq:manning}
\end{equation}
where $\sigma_\text{LJ,bond}$ is the average bond length, and $l_B$ is the Bjerrum length. Manning's theory is known to provide good predictions of the effective charge in the limit of long chains, but fails for very short chains~\cite{grass09a,frank09a,hickey12a}.

\begin{figure}[!htb]
	\includegraphics[width=0.9\linewidth]{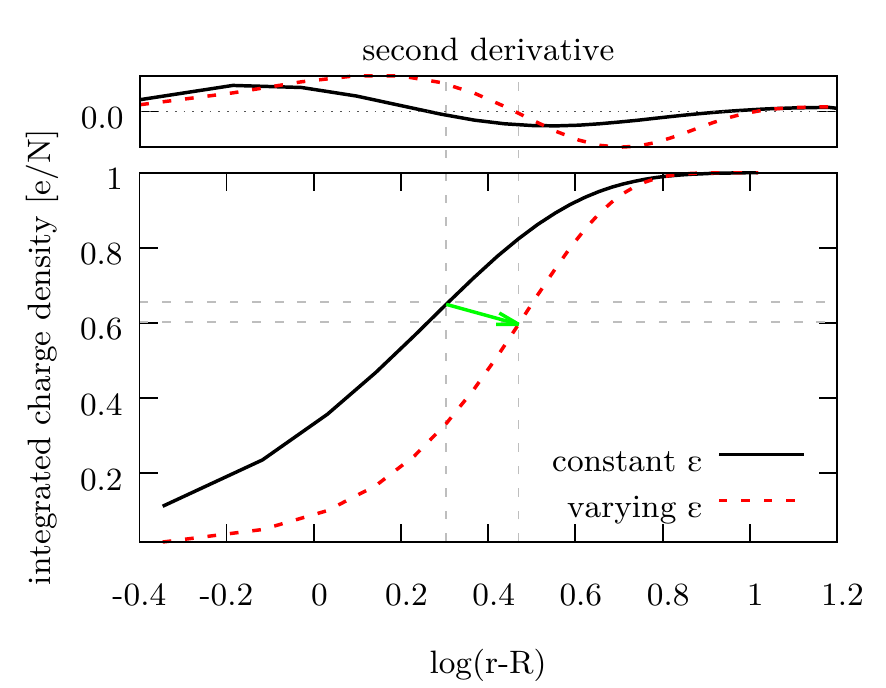}
	\caption{Normalized integrated charge density around the polyelectrolyte, plotted against the logarithm of the distance. The inflection point of the graph, pointed out by gray dashed lines, can be determined via a second derivative (upper plot, the dotted line marks zero) and is seen as the boundary between condensed and free counterions. For varying permittivity, this inflection point moves (green arrow) further away from the surface and to a lower integrated charge value, suggesting a higher effective charge and mobility of the polyelectrolyte.}
	\label{fig:integrated-charge-density}
\end{figure}

Regardless of the chain length one can calculate the (static) effective charge directly using the method developed by Belloni~\cite{belloni98a} and Deserno~\cite{deserno00a}. We plot the integrated charge density as a function of the logarithm of the distance to the backbone in Fig.~\ref{fig:integrated-charge-density}. The counterions closer to the backbone than the inflection point of this graph are considered condensed~\cite{deserno00a,grass08a}.

\begin{figure}[htbp]
	\subfloat[dynamic effective charge\label{fig:effective-charge-langevin}]{%
		\includegraphics[width=0.9\linewidth]{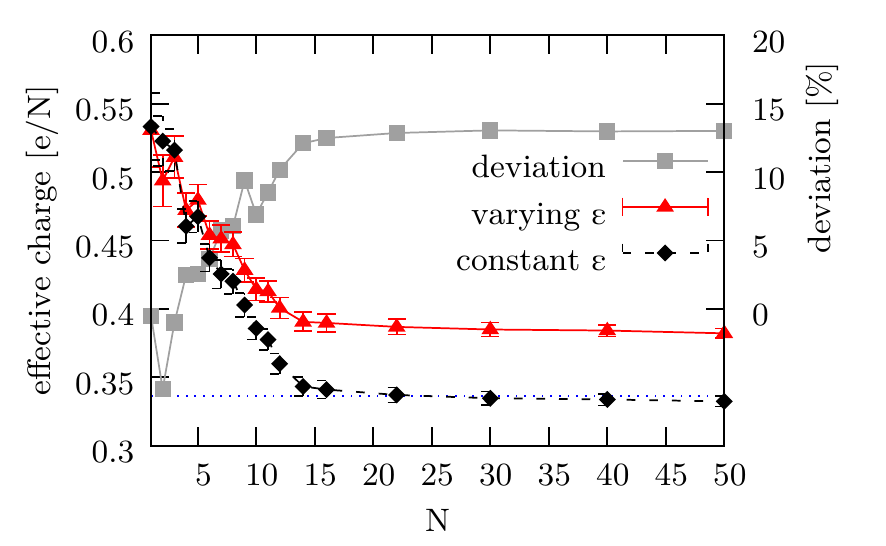}
	}\\
	\subfloat[static effective charge\label{fig:effective-charge-static}]{%
		\includegraphics[width=0.9\linewidth]{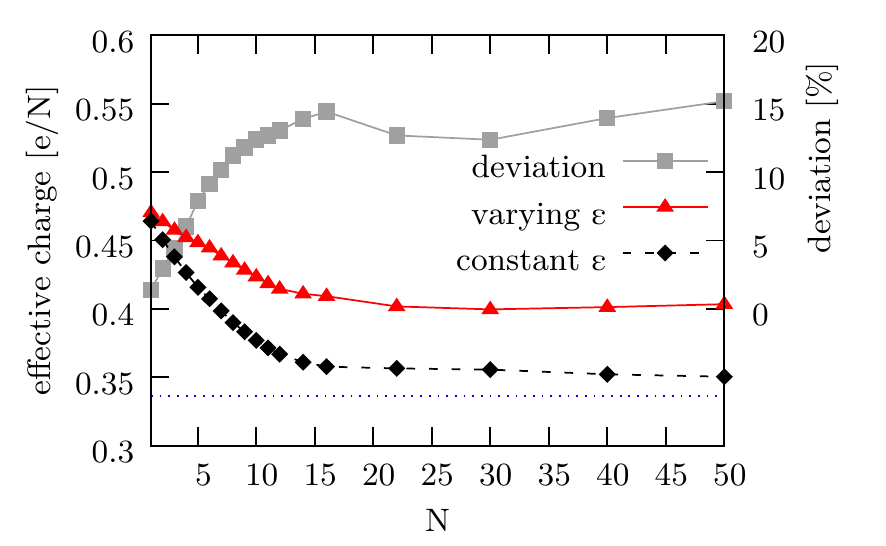}
	}
	\caption{Effective charge of the polyelectrolyte calculated in three different ways. The dashed line in both graphs depicts the effective charge predicted by Manning~\cite{manning69a}. (a) shows the dynamic effective charge calculated via the measured mobility with an applied electric field and without hydrodynamic interaction (Langevin dynamics). (b) shows the static effective charge calculated from the integrated charge density as sketched in Fig.~\ref{fig:integrated-charge-density}. The simulations with (red triangles) and without (black diamonds) varying dielectric permittivity show very little deviation (gray squares, right axis) at shorter polymer lengths and the difference increases until $N\approx 15$.}
	\label{fig:effective-charges}
\end{figure}

In Fig.~\ref{fig:integrated-charge-density} we see that including variations in the local permittivity causes the inflection point to move (green arrow) to longer distances $r$ from the backbone. At the same time, the number of condensed counterions $N_\text{cci}$ decreases, increasing the effective charge $Q_\text{eff} = (N-N_\text{cci})$, resulting in an increase electrophoretic mobility of the polyelectrolyte. 

In the case of Langevin dynamics simulations, one can calculate the effective friction $\Gamma_\text{eff}$ directly, resulting in:
\begin{equation}
	\mu =  \frac{Q_\text{eff}}{\Gamma_\text{eff}} \approx \frac{Q_\text{eff}}{N\Gamma},
\end{equation}
where $N$ is the number of monomers, and $\Gamma=1 m_0/\tau$ is the friction coefficient of an individual MD bead. A rescaled version of the mobility, $Q_\text{eff}/N = \Gamma \mu$, is plotted in Fig.~\ref{fig:effective-charge-langevin} and matches Manning's prediction in the long chain limit.

Both the results for the Langevin simulations and the static effective charge calculations in Fig.~\ref{fig:effective-charges} yield very similar results, with a difference of less than 3\% in the free draining limit. For a constant dielectric background, the effective charge for polyelectrolytes longer than $N=15$ monomers fits well with Manning's prediction~\cite{manning69a}. Our simulations with a flexible varying dielectric permittivity show a significant change compared to our simulations with constant permittivity in effective charge, with the difference being roughly 15\% for long chains. This explains why we see an approximately 15\% increase in the electrophoretic mobility when taking into account variations in the local dielectric constant.

\begin{changed}
We should add that taking into account only an increase in ionic attraction due the lower dielectric constant in the vicinity of the polyelectrolyte, as done for example in Ref.~\cite{muthukumar04a}, typically leads to a stronger electrostatic attraction, and therefore to a decrease in the effective charge of the polyelectrolyte. However, in our simulations the dielectric permittivity changes gradually in the vicinity of the polyelectrolyte, which results in a solvation force that pushes counterions further away from the backbone. This leads to a significantly widened electric double layer, as shown previously in Ref. ~\cite{fahrenberger14a} and in section IV,  and therefore to less tightly bound counterions and an increased effective charge.
\end{changed}

We have shown that the inclusion of varying dielectric permittivity  leads to a significant increase in the electrophoretic mobility as well as the number of free counterions, since fewer counterions are condensed to the polymer backbone. These two effects should combine in an additive way when looking at the conductivity. In our simulations including an external electric field, we measured the conductivity $\sigma$ using the relation:
\begin{equation}
 \sigma = \frac{\Vect{J}}{\Vect{E}},
\end{equation}
where $\Vect{J}$ is the current density, and $\Vect{E}$ is the applied external electric field.

\begin{figure}[htbp]
	\includegraphics[width=0.9\linewidth]{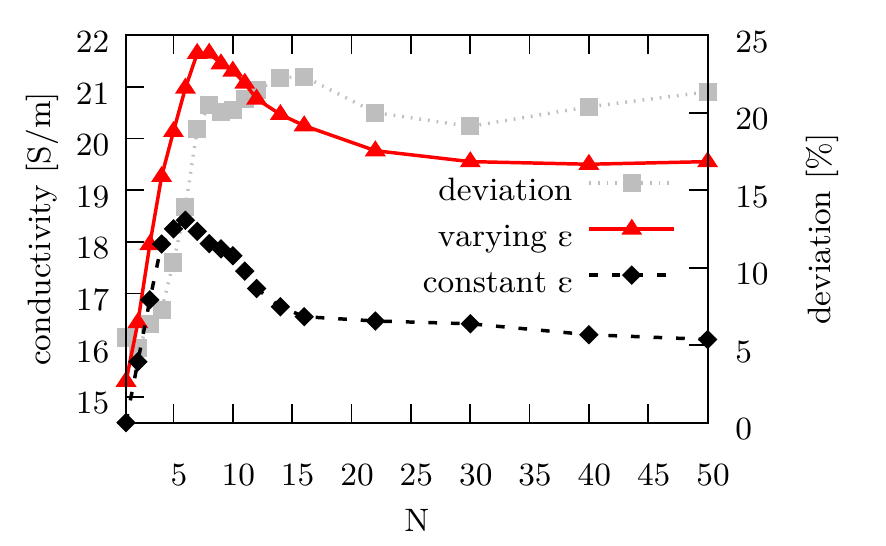}
	\caption{The conductivity as a function of the polymer length $N$. The deviation between simulations with and without varying permittivity is even more pronounced, since the higher mobility is accompanied by more uncondensed charges that can contribute to the conductivity.}
	\label{fig:conductivity}
\end{figure}

\begin{figure}[htbp]
	\includegraphics[width=0.9\linewidth]{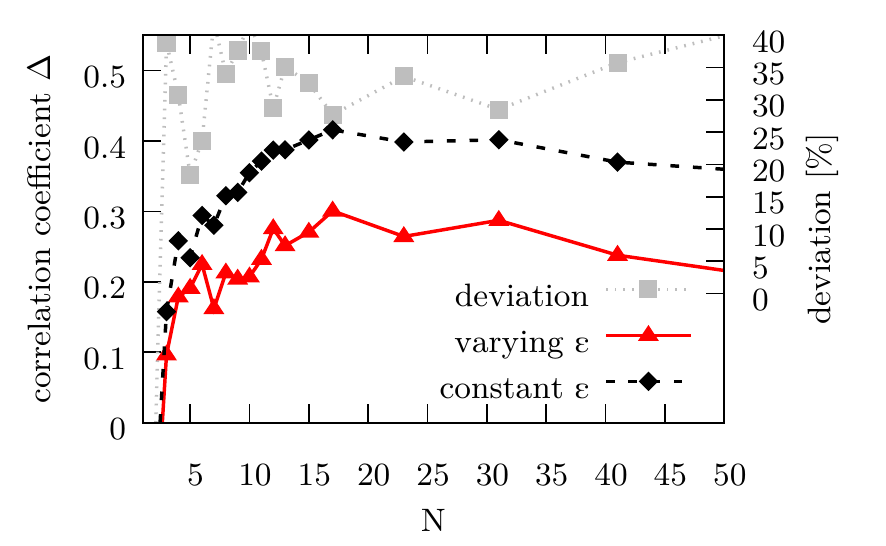}
	\caption{The correlation coefficient shows a significant decrease for the simulation with varying dielectric permittivity. This is due to the weaker coupling between the polyelectrolyte and the counterions, and should be observable in experiment.}
	\label{fig:correlation-coefficient}
\end{figure}

The conductivities with and without variation in the permittivity were converted to SI units using the relation
\begin{equation}
 \sigma^\text{SI} = \sigma^\text{MD} \frac{(e^\text{SI})^2 (\sigma_\text{LJ}^\text{MD})^3\zeta^\text{SI}}{(e^\text{MD})^2(\sigma_\text{LJ}^\text{SI})^3\zeta^\text{MD}},
\end{equation}
where $\sigma_\text{LJ}$ is the length scale, $e$ is the unit charge and $\zeta$ is the friction coefficient of an ion. The superscript denotes the according units system. To calculate the friction coefficient of an ion in SI units we used the Stokes relation $\zeta^\text{SI}=6\pi \eta \sigma^\text{SI}$, where
 $\eta=\SI{8.9e-4}{Pa.s}$ is the dynamic viscosity of water.
For the friction coefficient $\zeta^\text{MD}$ we used the expression from Ahlrichs and D\"unweg~\cite{ahlrichs99a}
\begin{equation}
	\zeta^\text{MD} = \frac{1}{\Gamma} = \frac{1}{\Gamma_0} + \frac{1}{25\eta a},
\end{equation}
where $\Gamma_0$ (the bare friction), $\eta$ (the dynamic viscosity), and $a$ (the grid spacing) are the lattice-Boltzmann parameters. The results are shown in Fig.~\ref{fig:conductivity} and are very comparable for short polymer lengths, but for $N>12$ taking into account variations in the local dielectric constant results in a 20\% increase in the conductivity. 
To relate this to experimentally accessible results, we are interested in the ratio between the measured conductivity $\sigma^{M}$ and the ideal conductivity $\sigma^{id}$. The ideal conductivity simply assumes that the mobility of all charge carriers is $\mu=q/6\pi \eta R_\text{H}$, which essentially ignores the hydrodynamic and electrostatic coupling between charge carriers. This lets us define a correlation coefficient
\begin{equation}
	\Delta = 1 - \frac{\sigma^{M}}{\sigma^{id}},
\end{equation}
which is $0$ when there is neither hydrodynamic nor electrostatic interactions between the charge carriers. The correlation coefficient quantifies the amount of friction and coupling between the polyelectrolyte and the counterions. The ideal conductivity can be determined with the diffusion constants of both species, via the Stokes-Einstein equation
\begin{equation}
	\sigma^{id} = \frac{Ne^2D^+}{k_\text{B}T}+\frac{N^2e^2D^-}{k_\text{B}T},
\end{equation}
where we can measure the diffusion constant for a single particle $D^{0}$ and determine the diffusion constant of the polyelectrolyte with
\begin{equation}
	D^- = \frac{D^{0}}{N} + \frac{1}{6\pi\eta R_H}\quad .
\end{equation}
The hydrodynamic radius $R_H$ in this equation is defined as
\begin{equation}
	\frac{1}{R_H} = \frac{1}{N^2}\left< \sum_{i\neq j}\frac{1}{r_{ij}} \right>\quad ,
\end{equation}
which we calculated directly from our simulations. The results for the correlation coefficient $\Delta$ are plotted in Fig.~\ref{fig:correlation-coefficient}.

We observe a significant difference between the approach with and without dielectric variations, exceeding 30\% for longer polyelectrolytes. This clearly shows that the coupling between the polyelectrolyte chain and the surrounding counterions is significantly reduced when variations in the permittivity are taken into account. The difference is the result of the structural differences in the EDL. This setup should be more readily accessible experimentally, and allow for a quantitative comparison to our simulations.

\subsection{Conductivity for varying monomer concentration}

We now turn to the conductivity of polyelectrolytes in salt-free aqueous solution. In Fig.~\ref{fig:conductivity}, the conductivity plateaus at polymer lengths above 20 monomers. With the same simulation setup as before, we measured the conductivity for different monomer concentrations by keeping the box size constant at $(\SI{32}{nm})^3$ and adding polyelectrolyte chains and counterions. We calculated a mean value and standard deviation for the conductivity with lengths $N=30$, $45$, and $60$ to make sure that the results are independent of $N$. Renormalizing this data with the monomer concentration gives an equivalent conductivity per monomer $\Lambda$. We also normalize all the data to an extrapolated value of $\Lambda(C=0)=1$. Through this rescaling $\Lambda$ is independent of both the friction coefficient of the MD beads and the viscosity of the fluid. 

\begin{figure}[!htb]
	\centering
	\includegraphics[width=0.9\linewidth]{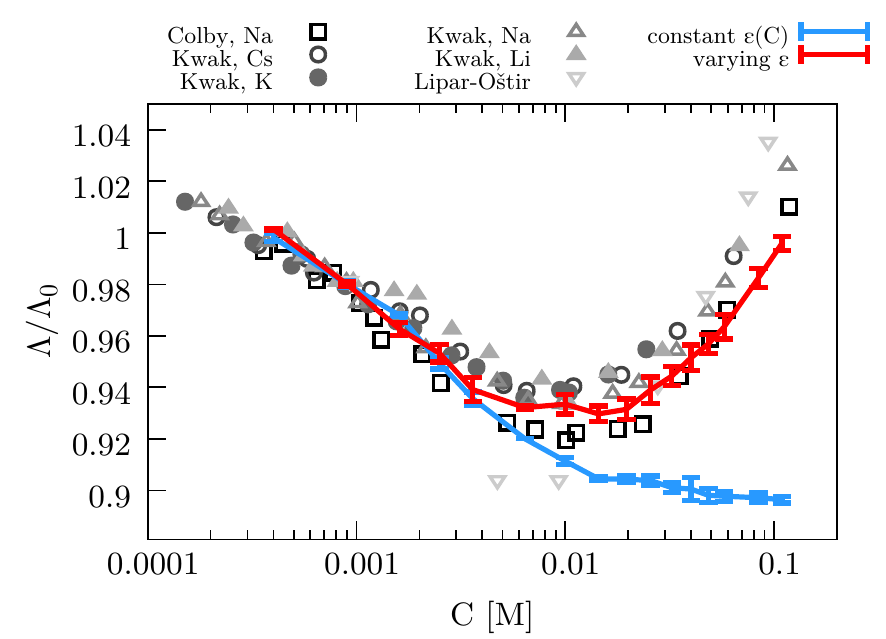}
	\caption{The equivalent conductivities $\Lambda/\Lambda_0$ as a function of the molar monomer concentration $C$. The experimentally observed minimum (gray symbols) is reproduced in simulations with varying $\varepsilon(\Vect{r})$ (red line), while simulations with constant but scaled $\varepsilon(C)$  (blue line) exhibit a different qualitative behavior.}
	\label{fig:conductivity-concentration}
\end{figure}

\begin{figure}[!htb]
	\centering
	\includegraphics[width=\linewidth]{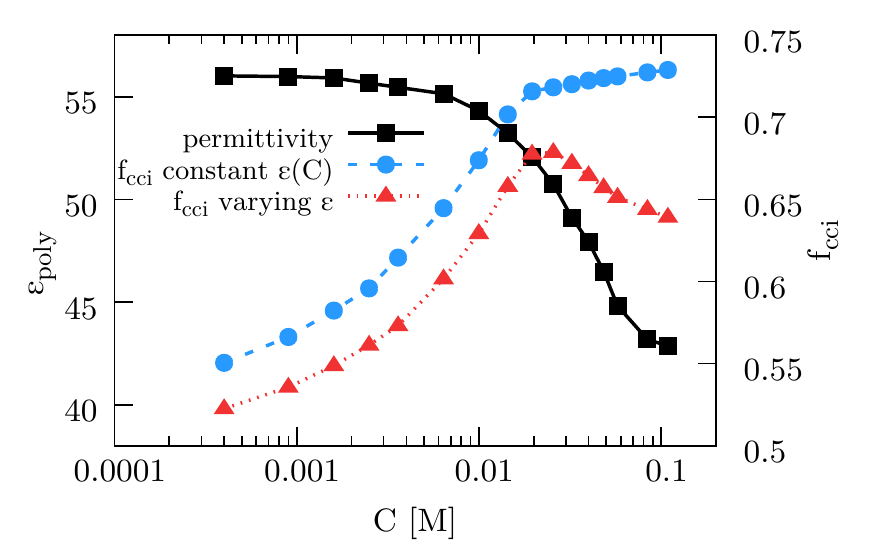}
	\caption{Average permittivity in the vicinity of the polyelectrolytes $\varepsilon_\text{poly}$ as calculated by the adaptive scheme (black squares, left axis), and the fraction of condensed counterions $f_\text{cci}$ (right axis) for constant but scaled $\varepsilon(C)$ (blue circles) and varying $\varepsilon(\Vect{r})$ (red triangles). $f_\text{cci}$ has a maximum for variable permittivity around the conductivity minimum observed before. The salt dependent permittivity keeps decreasing even though the number of condensed counterions almost stagnates at higher concentrations, which is due to the coiling of the polyelectrolytes.}
	\label{fig:epsilon-cci}
\end{figure}

Let us first consider the case of constant background permittivity, the blue line in Fig.~\ref{fig:conductivity-concentration}. Note that we set the background permittivity using Eq.~\ref{eq:salt-map}, although the results are almost identical to simply using a constant permittivity of $\varepsilon=78.5$ for all monomer concentrations~\cite{fahrenberger15b}.
The equivalent conductivity drops dramatically until a monomer concentration $C \approx \SI{0.01}{M}$.
In this regime, the Debye layer shrinks leading to more and more counterions condensing on the polyelectrolyte backbone as seen in Fig.~\ref{fig:epsilon-cci}. At higher monomer concentrations $C > \SI{0.01}{M}$, both the equivalent conductivity and fraction of condensed counterions continue to increase, but at a much slower rate.

\begin{figure}[!htb]
	\subfloat[$C=\SI{0.001}{M}$]{%
		\includegraphics[angle=90,width=0.35\linewidth]{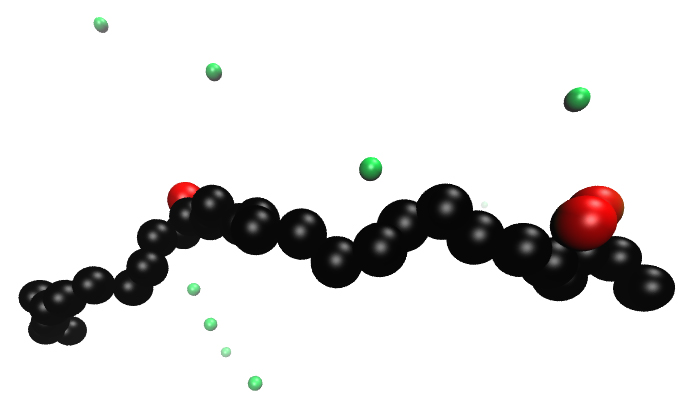}
	}
	\subfloat[$C=\SI{0.01}{M}$]{%
		\includegraphics[width=0.3\linewidth]{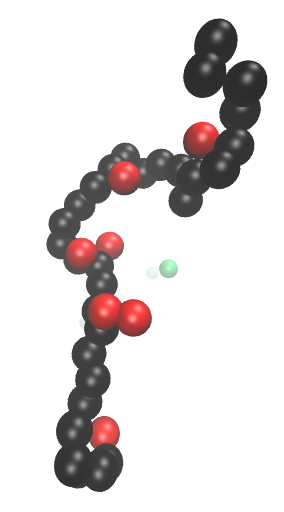}
	}
	\subfloat[$C=\SI{0.1}{M}$]{%
		\raisebox{7mm}{\includegraphics[width=0.35\linewidth]{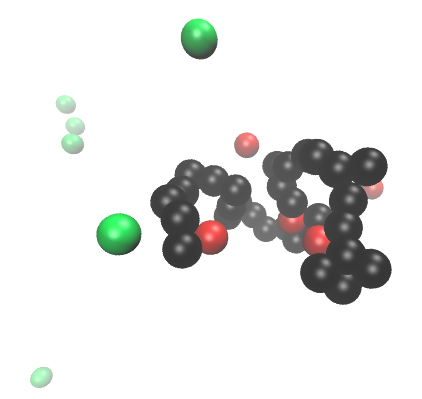}}
	}
	\caption{Snapshots from our simulations. (a)~At low monomer concentration, the polyelectrolyte has few condensed counterions (red) and is stretched out because of electrostatic repulsion of the backbone. (b)~At higher monomer concentration, the number of condensed counterions is significantly increased. (c)~At high monomer concentration, the polyelectrolyte coils. In the simulations with varying dielectric permittivity depending on the local charge concentration, this coiling of highly charged monomers leads to a lower average permittivity around the polymer backbone.}
	\label{fig:stretched-vs-coiled}
\end{figure}

\begin{figure}[!htb]
	\centering
	\includegraphics[width=\linewidth]{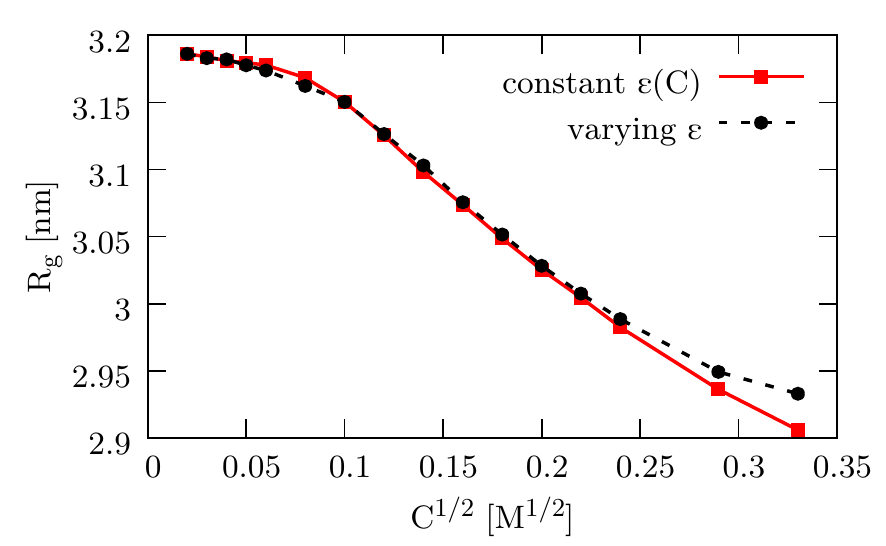}
	\caption{Average radius of gyration $R_\text{G}$ of the polyelectrolytes as a function of the polyelectrolyte concentration for a uniform background permittivity $\varepsilon(C)$ (red squares, solid line) and a locally varying permittivity $\varepsilon(\Vect{r})$ (black circles, dashed line).}
	\label{fig:rg-vs-conc}
\end{figure}

Simulations including variations in the local permittivity also display an initial decrease of the conductivity with increasing polyelectrolyte concentrations, which is again due to an increase in the fraction of condensed counterions in Fig.~\ref{fig:epsilon-cci}.
There is a clear minimum around $C =\SI{0.01}{M}$, after which the equivalent conductivity begins to rise. This can be attributed to the decrease in
the fraction of condensed counterions in Fig.~\ref{fig:epsilon-cci}. The decrease in the fraction of condensed counterions can in turn be related to the large decrease in the local relative permittivity from around 55 in the dilute limit, to approximately 41 at the highest monomer concentrations (the red line in Fig.~\ref{fig:epsilon-cci}). This must be the result of an increase in the local ion concentration, since it alone determines the permittivity in our simulations via equation~\ref{eq:salt-map}. The reason for the increased ion concentration is that the decreasing Debye length causes the polymers to coil significantly as can be seen in figures~\ref{fig:stretched-vs-coiled} and \ref{fig:rg-vs-conc}.

We also compare our results in Fig.~\ref{fig:conductivity} to existing experimental data from Kwak and Hayes~\cite{kwak75a}, Colby~\etal{}~\cite{colby97a}, and Lipar-O\v{s}tir~\etal{}~\cite{liparostir09a}. Since the raw results greatly depend on the hydrodynamic radius of the solvated counterions and differ by more than a factor of $2$, we normalize all the data to an extrapolated value of $\Lambda(C=0)=1$. The experimental data show excellent agreement with our simulation results. The slight differences are most likely due to enthalpic factors between the counterions and the polyelectrolytes or the specific structure of the polyelectrolyte, however, these are clearly secondary factors.

\section{Conclusions}

We used our novel implementation of a local electrostatic method to include spatial and temporal changes in the local dielectric permittivity. We then performed simulations of polyelectrolytes solutions. We first looked at a fixed charged straight polymer, and found excellent agreement of the counterion distribution with an additional iterative method. Overall, the results were quite similar to findings for the counterion distribution around colloidal particles~\cite{fahrenberger13c,ma14a}. We have also showed that the inclusion of a dielectric jump in the dielectric permittivity at the surface of the polyelectrolyte was insufficient to even qualitatively reproduce the same counterion distribution~\cite{fahrenberger15b}. This is because the dielectric jump fails to take into account the force caused by the gradient in the dielectric permittivity related to the solvation energy of the ions. \begin{changed}Our results show that this force can not be neglected in this setup, since it significantly widens the electric double layer and therefore reduces counterion condensation. While many studies have looked at the effect of reduced permittivity within a polymer, colloid, or surface, our results clearly show that the gradient in the dielectric constant adjacent to the interface actually plays a much larger role.\end{changed}

We then applied our algorithm to a free, flexible polyelectrolyte. Interestingly, the distribution of counterions around the backbone of the fluctuating polymer are almost identical to the results for the fixed, straight polymer. Despite a qualitatively different structure of the electric double layer around the charged polymer, we observed almost no change in the radius of gyration of the chains when taking into account salt-dependent variations in the local dielectric constant. This demonstrates that the change in the local counterion distribution only has a minor influence on the conformational properties of the polyelectrolyte.

Next, we looked at the role of varying permittivity on the mobility of a polyelectrolyte in aqueous solution subject to an external electric field. We found that the qualitative behavior of earlier simulations~\cite{grass08a,frank09a} is reproduced, in agreement with experimental data~\cite{grass09a,hoagland99a}. That being said, the difference in the distribution of counterions within the electric double layer had a significant quantitative influence on the electrophoretic mobility. The mobility and effective charge of the polyelectrolytes, in comparison to a simulation without varying permittivity, shows an average increase of around 9\% for long polyelectrolytes. We observed a roughly 20\% increase in the electric conductivity when taking into account changes in the local permittivity, since both the increased effective charge of the polyelectrolyte and the increased number of uncondensed counterions increase by approximately 10\%.

We then investigated the role of monomer concentration on the conductivity of salt-free polyelectrolyte solutions. There was a clear qualitative difference between the simulations with constant dielectric background and the ones using our adaptive scheme for variations in the permittivity, where the latter does not show a continuous decrease of conductivity with rising monomer concentration, but exhibits a distinct minimum and subsequent rise in the equivalent conductivity. The same nonmonotonic behavior has previously been observed in experiments~\cite{kwak75a,colby97a,liparostir09a}, and our simulation results closely match experiment and even exhibit the conductivity minimum at the same monomer concentration.

Our results can be succinctly summarized by saying that the varying dielectric permittivity causes a thickening of the Debye layer, which has only a minor influence on the conformation of the chain. Despite not making dramatic changes to the static properties of the system, the electrophoretic mobility and electric conductivity is significantly larger. Experimental data for the equivalent conductivity at increasing monomer concentration is reproduced using our scheme for varying permittivity, while the simulations assuming a constant dielectric background show qualitatively different behavior. This indicates that correctly accounting for variations in the local permittivity due to the local salt concentration is necessary to achieve quantitative and qualitative agreement with experiment. Most likely taking into account local variations in the dielectric permittivity would lead to similar quantitative differences for other systems such as the electrophoretic mobility of colloids, and electroosmotic flow in microfluidics.

\section*{Acknowledgements}

We thank the DFG for support through the project HO/1108-22, \begin{changed}the  SFB 716 TP C5 , and the Cluster of Excellence in Simulation Technology (EXC 310) at the University of Stuttgart\end{changed}. Furthermore we acknowledge partial funding from the German Ministry of Science and Education (BMBF) under grant 01IH08001.


%

\end{document}